\documentstyle[12pt]{article}

\def\noi{\noindent}

\def\nqq{\hspace{-2em}}

\def\barr{\left(\begin{array}}
\def\earr{\end{array}\right)}
\def\beq#1{\begin{equation}\label{#1}}
\def\eeq{\end{equation}}
\def\ber#1{\begin{eqnarray}\label{#1} &&\nqq}
\def\eer{\end{eqnarray}}
\def\eern{\nonumber \end{eqnarray}}
\def\nn{\nonumber\\ &&\nqq}
\def\mm{\\ &&\nqq}

\catcode`\@=11 \@addtoreset{equation}{section}\catcode`\@=12

\newcommand{\R}{\mbox{\bf R}}
\newcommand{\Z}{\mbox{\bf Z}}
\newcommand{\N}{\mbox{\bf N}}
\newcommand{\SL}{\mathop{\rm sl}\nolimits}
\newcommand{\SO}{\mathop{\rm so}\nolimits}
\newcommand{\SP}{\mathop{\rm sp}\nolimits}

\newcommand{\diag}{\mathop{\rm diag}\nolimits}
\newcommand{\sign}{\mathop{\rm sign}\nolimits}

\newcommand{\eps}{\varepsilon}
\newcommand{\tri}{\triangle}

\newcommand{\e}[1]{\mathop{\rm e}\nolimits^{#1}}

\newcommand{\p}{\partial}

\newcommand{\ds}{\displaystyle}

\newcommand{\fnm}{\footnotemark}
\newcommand{\fnt}{\footnotetext}

\begin{document}

\thispagestyle{empty}

\vspace*{2cm}

\begin{center}
\Large\bf
SIGMA-MODEL SOLUTIONS AND \\ INTERSECTING P-BRANES RELATED \\
TO LIE ALGEBRAS \\[20pt]

\large\bf
M.A. Grebeniuk\fnm[1]\fnt[1]{mag@gravi.phys.msu.su} \\[15pt]

\it
Moscow State University, Physical Faculty, Department of \\
Theoretical Physics, Moscow 117234, Russia \\[20pt]

\bf
and V.D. Ivashchuk\fnm[2]\fnt[2]{ivas@rgs.phys.msu.su} \\[15pt]

\it
Center for Gravitation and Fundamental Metrology, VNIIMS \\
3-1, M.Ulyanovoy Str., Moscow 117313, Russia
\end{center}

\vspace{2cm}

\noi
A family of Majumdar-Papapetrou type solutions in $\sigma$-model of
$p$-brane origin is obtained for all direct sums of finite-dimensional
simple Lie algebras. Several examples of $p$-brane dyonic configurations
in $D=10$ (IIA) and $D=11$ supergravities corresponding to the Lie
algebra $A_2$ are considered.

\newpage

\section{Introduction}

In this paper we give explicit relations for the Majumdar-Papapetrou
type solutions from \cite{IM5}, corresponding to direct sums of
simple finite-dimensional Lie algebras (e.g. to semisimple Lie algebras).
These solutions appear in the $\sigma$-model of $p$-brane
origin \cite{IM4}--\cite{IMR} and are governed by a set of harmonic
functions.

For $p$-brane configurations the obtained solutions describe bound states
of $p$-brane extremal configurations with non-standard intersection rules
(for non-extremal case see \cite{IMJ}). The standard "orthogonal" solutions
(see \cite{IM4}--\cite{AEH} and refs. therein) correspond to the
Lie algebras $A_1\oplus\dots\oplus A_1$.

Here we also consider several examples of the solutions in $D=10$ IIA
supergravity and $D=11$ supergravity. These dyonic solutions correpond to
the Lie algebra $A_2=\SL(3)$ and may be interesting as possible
applications in MQCD \cite{EGK}--\cite{BIKSY}.

\section{Majumdar-Papapetrou solutions in $\sigma$-model}

We consider the $\sigma$-model governed by the action
\ber{2.1}
S_\sigma=\int d^{d_0}x\sqrt{|g^0|}\biggl\{R[g^0]-\hat G_{AB}
g^{0\mu\nu}\p_\mu\sigma^A\p_\nu\sigma^B \nn
-\sum_{s\in S}\eps_s\e{-2U_A^s\sigma^A}
g^{0\mu\nu} \p_\mu\Phi^s\p_\nu\Phi^s\biggr\},
\eer
where $g^0=g^0_{\mu\nu}(x)dx^\mu\otimes dx^\nu$ is a metric on
$d_0$-dimensional manifold $M_0$, $\sigma=(\sigma^A)\in\R^N$ and
$\Phi=(\Phi^s,s\in S)$ is a set of  scalar fields ($S\ne\emptyset$),
$(\hat G_{AB})$ is a non-degenerate symmetric matrix, $U^s=(U_A^s)\in\R^N$
are vectors and $\eps_s=\pm1$, $s\in S$. The scalar product is defined
as follows
\ber{2.2}
(U,U')=\hat G^{AB}U_AU'_B,
\eer
for $U,U'\in\R^N$, where $(\hat G^{AB})=(\hat G_{AB})^{-1}$.

Let the set $S$ be a union of $k$ non-intersecting (non-empty) subsets
$S_1,\dots,S_k$:
\ber{2.3}
S=S_1\sqcup\dots\sqcup S_k,
\eer
$S_i\ne\emptyset$, $i=1,\dots,k$, and
\ber{2.4}
(U^s,U^{s'})=0
\eer
for all $s\in S_i$, $s'\in S_j$, $i\ne j$; $i,j=1,\dots,k$. According to
(\ref{2.4}) the set of vectors $(U^s,s\in S)$ has a block-orthogonal
structure with respect to the scalar product (\ref{2.2}), i.e. it is
splitted into $k$ mutually-orthogonal blocks $(U^s,s\in S_i)$,
$i=1,\dots,k$.

Equations of motion corresponding to (\ref{2.1}) have the following
form
\ber{2.5}
R_{\mu\nu}[g^0]=\hat G_{AB}\p_\mu\sigma^A\p_\nu\sigma^B+
\sum_{s\in S}\eps_s\e{-2U_A^s\sigma^A}\p_\mu\Phi^s\p_\nu\Phi^s, \mm
\label{2.6}
\hat G_{AB}\tri[g^0]\sigma^B+\sum_{s\in S}\eps_sU_A^s\e{-2U_C^s\sigma^C}
g^{0\mu\nu}\p_\mu\Phi^s\p_\nu\Phi^s =0, \mm
\label{2.7}
\p_\mu\left(\sqrt{|g^0|}g^{0\mu\nu}\e{-2U_A^s\sigma^A}
\p_\nu\Phi^s\right)=0,
\eer
$s\in S$. Here  $\tri[g^0]$ is the Laplace-Beltrami operator
corresponding to $g^0$.

{\bf Proposition 1 \cite{IM5}.} Let $(M_0,g^0)$ be Ricci-flat
\ber{2.8}
R_{\mu\nu}[g^0]=0.
\eer
Then the field configuration
\ber{2.9}
g^0, \qquad \sigma^A=\sum_{s\in S}\eps_sU^{sA}\nu_s^2\ln H_s, \qquad
\Phi^s=\frac{\nu_s}{H_s},
\eer
$s\in S$, satisfies the field equations (\ref{2.5})--(\ref{2.7}) with
$V=0$ if (real) numbers $\nu_s$ obey the relations
\ber{2.10}
\sum_{s'\in S}(U^s,U^{s'})\eps_{s'}\nu_{s'}^2=-1
\eer
$s\in S$, functions $H_s >0$ are harmonic, i.e. $\tri[g^0]H_s=0$, $s\in S$,
and $H_s$ are coinciding inside blocks: $H_s=H_{s'}$ for $s,s'\in S_i$,
$i=1,\dots,k$.

In special orthogonal case, when any block contains only one vector (i.e.
all $|S_i|=1$) the Proposition 1 coincides with that of \cite{IMC}. In a
general case vectors inside each block $S_i$ are not orthogonal. The
solution under consideration depends on $k$ independent harmonic functions
corresponding to $k$ blocks. For a given set of vectors $(U^s,s\in S)$ the
maximal number $k$ arises for irreducible block-orthogonal decomposition
(\ref{2.3}), (\ref{2.4}), when any block $(U^s,s\in S_i)$ can not be
splitted into two mutually-orthogonal subblocks.

\section{Solutions related to Lie algebras}

Here we put
\ber{3.1}
(U^s,U^s)\ne0
\eer
for all $s\in S$ and introduce the quasi-Cartan matrix $A=(A^{ss'})$
\ber{3.2}
A^{ss'} \equiv \frac{2(U^s,U^{s'})}{(U^{s'},U^{s'})},
\eer
$s,s'\in S$. From (\ref{2.4}) we get a block-diagonal structure
of $A$:
\ber{3.3}
A=\diag(A_{(1)},\dots,A_{(k)}),
\eer
where $A_{(i)}=(A^{ss'},s,s'\in S_i)$, $i=1,\dots,k$. Here the set $S$ is
ordered, $S_1<\dots<S_k$, and the order in $S_i$ is inherited by the order
in $S$.

For $\det A_{(i)}\ne0$ relation (\ref{2.10}) may be rewritten in the
equivalent form
\ber{3.4}
\eps_s\nu_s^2(U^s,U^s)=-2\sum_{s'\in S}A_{ss'}^{(i)},
\eer
$s\in S_i$, where $(A_{ss'}^{(i)})=A_{(i)}^{-1}$. Thus, eq. (\ref{2.10})
may be resolved in terms of $\nu_s$ for certain $\eps_s=\pm1$, $s\in S_i$.

Let $A$ be a generalized Cartan matrix \cite{Kac,FS}. In this case
\ber{3.6}
A^{ss'}\in -\Z_+ \equiv\{0,-1,-2,\dots\}
\eer
for $s \ne s'$ and $A$ generates generalized symmetrizable Kac-Moody
algebra \cite{Kac,FS}. Here there are three possibilities for $A_{(i)}$:
a) $\det A_{(i)}>0$, b) $\det A_{(i)}<0$ and c) $\det A_{(i)}=0$. For
$\det A_{(i)}\ne0$ the corresponding Kac-Moody algebra is simple, since
$A_{(i)}$ is indecomposable \cite{FS}. In this paper we consider only
the case $\det A_{(i)}\ne0$.

\subsection{Simple finite-dimensional Lie algebras}

Let $A_{(i)}$ be a Cartan matrix of a simple finite-dimensional Lie algebra
$L_i$ and hence $\det A_{(i)} > 0$, $A_{(i)}^{ss'}\in\{0,-1,-2,-3\}$,
$s\ne s'$, $s\in S_i$, $i = 1, \ldots, k$. In this case the matrix $A$
from (\ref{3.3}) is the Cartan matrix for the Lie algebra
\ber{3.8a}
L = \oplus \sum_{j=1}^{k} L_j.
\eer
When $\dim L_j \neq 1$, for all $j = 1, \ldots, k$,
$L$ is semisimple.

The elements of inverse matrix $A_{(i)}^{-1}$ are positive (see Ch.7 in
\cite{FS}) and hence we get from (\ref{3.4})
\ber{3.8}
\eps_s(U^s,U^s)<0,
\eer
$s\in S_i$.

In summary \cite{FS}, there are four infinite series of simple Lie algebras,
which are denoted by
\ber{3.9}
A_r \ (r\ge1), \quad B_r \ (r\ge3), \quad
C_r \ (r\ge2), \quad D_r \ (r\ge4),
\eer
and in addition five isolated cases, which are called
\ber{3.10}
E_6, \quad E_7, \quad E_8, \quad G_2, \quad F_4.
\eer
In all cases the subscript denotes the rank of the algebra. The algebras
in the infinite series of simple Lie algebras are called the classical
(Lie) algebras. They are isomorphic to the matrix algebras
\ber{3.11}
A_r\cong\SL(r+1), \quad B_r\cong\SO(2r+1), \quad
C_r\cong\SP(r), \quad D_r\cong\SO(2r).
\eer
The five isolated cases are referred to as the exceptional Lie algebras.

For the simple Lie algebras of type $A_r$, $D_r$, $E_6$, $E_7$ and $E_8$,
all root have the same length, and any two nodes of Dynkin diagram are
connected by at most one line. In the other cases there are roots of two
different lengths, the length of the long roots being $\sqrt2$ times the
length of the short roots for $B_r$, $C_r$ and $F_4$, and $\sqrt3$ times
for $G_2$, respectively.

Now, let us consider the solutions in the sigma-model related to the simple
Lie algebras considered above.

{\bf $A_r$ series.} Let $A_{(i)}$ be $r\times r$ Cartan matrix for the Lie
algebra $A_r= \SL(r+1)$, $r\ge 1$. This matrix is described graphically by
the Dynkin diagram pictured on Fig.1.
\begin{center}
\bigskip
\begin{picture}(66,11)
\put(5,10){\circle*{2}}
\put(15,10){\circle*{2}}
\put(25,10){\circle*{2}}
\put(55,10){\circle*{2}}
\put(65,10){\circle*{2}}
\put(5,10){\line(1,0){30}}
\put(45,10){\line(1,0){10}}
\put(5,5){\makebox(0,0)[cc]{1}}
\put(15,5){\makebox(0,0)[cc]{2}}
\put(25,5){\makebox(0,0)[cc]{3}}
\put(40,10){\makebox(0,0)[cc]{$\dots$}}
\put(55,5){\makebox(0,0)[cc]{$r-1$}}
\put(65,5){\makebox(0,0)[cc]{$r$}}
\put(55,10){\line(1,0){10}}
\end{picture} \\[5pt]
\small
Fig.1. \it Dynkin diagram for $A_r$ Lie algebra
\end{center}
(For $s\ne s'$, $A_{(i)}^{ss'}=-1$ if nodes $s$ and $s'$ are connected by
a line on the diagram and $A_{(i)}^{ss'}=0$ otherwise). Using the relation
for the inverse matrix $A_{(i)}^{-1}=(A_{ss'}^{(i)})$ (see Sect.7.5 in
\cite{FS})
\ber{3.12}
A_{ss'}^{(i)}=\frac1{r+1}\min(s,s')[r+1-\max(s,s')]
\eer
we may rewrite (\ref{3.4}) as follows
\ber{3.13}
\eps_s\nu_s^2(U^s,U^s)=s(s-1-r),
\eer
$s\in\{1,\dots,r\}=S_i$.

{\bf $B_r$ and $C_r$ series.} Dynkin diagrams for these cases are
pictured on Fig.2.
\begin{center}
\bigskip
\begin{picture}(66,12)
\put(5,10){\circle*{2}}
\put(15,10){\circle*{2}}
\put(25,10){\circle*{2}}
\put(55,10){\circle*{2}}
\put(65,10){\circle*{2}}
\put(5,10){\line(1,0){30}}
\put(45,10){\line(1,0){10}}
\put(5,5){\makebox(0,0)[cc]{1}}
\put(15,5){\makebox(0,0)[cc]{2}}
\put(25,5){\makebox(0,0)[cc]{3}}
\put(40,10){\makebox(0,0)[cc]{$\dots$}}
\put(55,5){\makebox(0,0)[cc]{$r-1$}}
\put(65,5){\makebox(0,0)[cc]{$r$}}
\put(55,11){\line(1,0){10}}
\put(55,9){\line(1,0){10}}
\put(60,10){\makebox(0,0)[cc]{\large $>$}}
\end{picture} \qquad
\begin{picture}(66,12)
\put(5,10){\circle*{2}}
\put(15,10){\circle*{2}}
\put(25,10){\circle*{2}}
\put(55,10){\circle*{2}}
\put(65,10){\circle*{2}}
\put(5,10){\line(1,0){30}}
\put(45,10){\line(1,0){10}}
\put(5,5){\makebox(0,0)[cc]{1}}
\put(15,5){\makebox(0,0)[cc]{2}}
\put(25,5){\makebox(0,0)[cc]{3}}
\put(40,10){\makebox(0,0)[cc]{$\dots$}}
\put(55,5){\makebox(0,0)[cc]{$r-1$}}
\put(65,5){\makebox(0,0)[cc]{$r$}}
\put(55,11){\line(1,0){10}}
\put(55,9){\line(1,0){10}}
\put(60,10){\makebox(0,0)[cc]{\large $<$}}
\end{picture} \\[5pt]
\small
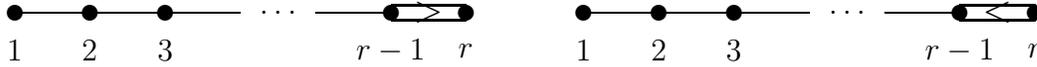
Fig.2. \it Dynkin diagrams for $B_r$ and $C_r$ Lie algebras
\end{center}
In these cases we have the following formulas for inverse matrices
$A_{(i)}^{-1}=(A_{ss'}^{(i)})$:
\ber{3.14}
A_{ss'}^{(i)}=\left\{\begin{array}{ll}
\min(s,s') & \mbox{for }s\ne r, \\[5pt]
\frac12s' & \mbox{for }s=r,
\end{array}\right. \quad
A_{ss'}^{(i)}=\left\{\begin{array}{ll}
\min(s,s') & \mbox{for }s'\ne r, \\[5pt]
\frac12s & \mbox{for }s'=r
\end{array}\right.
\eer
and relation (\ref{3.4}) takes the form
\ber{3.15}
\eps_s\nu_s^2(U^s,U^s)=\left\{\begin{array}{ll}
s(s-2r-1) & \mbox{for }s\ne r,\\
-\frac r2(r+1) & \mbox{for }s=r;
\end{array}\right.\quad
\eps_s\nu_s^2(U^s,U^s)=s(s-2r),
\eer
for $B_r$ and $C_r$ series respectively, $s\in\{1,\dots,r\}=S_i$.

{\bf $D_r$ series.} We have the following Dynkin diagram for this case
(Fig.3):
\begin{center}
\bigskip
\begin{picture}(63,31)
\put(5,20){\circle*{2}}
\put(15,20){\circle*{2}}
\put(25,20){\circle*{2}}
\put(55,20){\circle*{2}}
\put(5,20){\line(1,0){30}}
\put(45,20){\line(1,0){10}}
\put(5,15){\makebox(0,0)[cc]{1}}
\put(15,15){\makebox(0,0)[cc]{2}}
\put(25,15){\makebox(0,0)[cc]{3}}
\put(40,20){\makebox(0,0)[cc]{$\dots$}}
\put(55,15){\makebox(0,0)[rc]{$r-2$}}
\put(62,27){\circle*{2}}
\put(62,13){\circle*{2}}
\put(55,20){\line(1,1){7}}
\put(55,20){\line(1,-1){7}}
\put(62,31){\makebox(0,0)[cc]{$r$}}
\put(62,8){\makebox(0,0)[cc]{$r-1$}}
\end{picture} \\[5pt]
\small
Fig.3. \it Dynkin diagram for $D_r$ Lie algebra
\end{center}
and formula for the inverse matrix $A_{(i)}^{-1}=(A_{ss'}^{(i)})$ \cite{FS}:
\ber{3.16}
A_{ss'}^{(i)}=\left\{\begin{array}{ll}
\min(s,s') & \mbox{for }s,s'\notin\{r,r-1\}, \\[5pt]
\frac12s & \mbox{for }s\notin\{r,r-1\}, \ s'\in\{r,r-1\}, \\[5pt]
\frac12s' & \mbox{for }s\in\{r,r-1\}, \ s'\notin\{r,r-1\}, \\[5pt]
\frac14r & \mbox{for }s=s'=r \mbox{ or } s=s'=r-1, \\[5pt]
\frac14(r-2) & \mbox{for }s=r, \ s'=r-1 \mbox{ or vice versa.}
\end{array}\right.
\eer
The relation (\ref{3.4}) in this case reads
\ber{3.17}
\eps_s\nu_s^2(U^s,U^s)=\left\{\begin{array}{ll}
s(s-2r+1) & \mbox{for }s\notin\{r,r-1\},\\
-\frac r2(r-1) & \mbox{for }s\in\{r,r-1\},
\end{array}\right.
\eer
$s\in\{1,\dots,r\}=S_i$.

Let us consider the exceptional Lie algebras. Dynkin diagrams of
these algebras are pictured on Fig.4.
\begin{center}
\bigskip
\begin{picture}(46,21)
\put(5,10){\circle*{2}}
\put(15,10){\circle*{2}}
\put(25,10){\circle*{2}}
\put(35,10){\circle*{2}}
\put(45,10){\circle*{2}}
\put(25,20){\circle*{2}}
\put(5,10){\line(1,0){40}}
\put(25,10){\line(0,1){10}}
\put(5,5){\makebox(0,0)[cc]{1}}
\put(15,5){\makebox(0,0)[cc]{2}}
\put(25,5){\makebox(0,0)[cc]{3}}
\put(35,5){\makebox(0,0)[cc]{4}}
\put(45,5){\makebox(0,0)[cc]{5}}
\put(28,20){\makebox(0,0)[lc]{6}}
\end{picture} \qquad
\begin{picture}(56,21)
\put(5,10){\circle*{2}}
\put(15,10){\circle*{2}}
\put(25,10){\circle*{2}}
\put(35,10){\circle*{2}}
\put(45,10){\circle*{2}}
\put(25,20){\circle*{2}}
\put(5,10){\line(1,0){40}}
\put(25,10){\line(0,1){10}}
\put(5,5){\makebox(0,0)[cc]{1}}
\put(15,5){\makebox(0,0)[cc]{2}}
\put(25,5){\makebox(0,0)[cc]{3}}
\put(35,5){\makebox(0,0)[cc]{4}}
\put(45,5){\makebox(0,0)[cc]{5}}
\put(28,20){\makebox(0,0)[lc]{7}}
\put(55,10){\circle*{2}}
\put(45,10){\line(1,0){10}}
\put(55,5){\makebox(0,0)[cc]{6}}
\end{picture} \\[15pt]
\begin{picture}(66,21)
\put(5,10){\circle*{2}}
\put(15,10){\circle*{2}}
\put(25,10){\circle*{2}}
\put(35,10){\circle*{2}}
\put(45,10){\circle*{2}}
\put(45,20){\circle*{2}}
\put(5,10){\line(1,0){40}}
\put(45,10){\line(0,1){10}}
\put(5,5){\makebox(0,0)[cc]{1}}
\put(15,5){\makebox(0,0)[cc]{2}}
\put(25,5){\makebox(0,0)[cc]{3}}
\put(35,5){\makebox(0,0)[cc]{4}}
\put(45,5){\makebox(0,0)[cc]{5}}
\put(48,20){\makebox(0,0)[lc]{8}}
\put(55,10){\circle*{2}}
\put(65,10){\circle*{2}}
\put(45,10){\line(1,0){20}}
\put(55,5){\makebox(0,0)[cc]{6}}
\put(65,5){\makebox(0,0)[cc]{7}}
\end{picture} \qquad
\begin{picture}(36,12)
\put(5,10){\circle*{2}}
\put(15,10){\circle*{2}}
\put(25,10){\circle*{2}}
\put(35,10){\circle*{2}}
\put(5,10){\line(1,0){10}}
\put(15,11){\line(1,0){10}}
\put(15,9){\line(1,0){10}}
\put(25,10){\line(1,0){10}}
\put(20,10){\makebox(0,0)[cc]{\large $>$}}
\put(5,5){\makebox(0,0)[cc]{1}}
\put(15,5){\makebox(0,0)[cc]{2}}
\put(25,5){\makebox(0,0)[cc]{3}}
\put(35,5){\makebox(0,0)[cc]{4}}
\end{picture} \qquad
\begin{picture}(16,12)
\put(5,10){\circle*{2}}
\put(15,10){\circle*{2}}
\put(5,11){\line(1,0){10}}
\put(5,10){\line(1,0){10}}
\put(5,9){\line(1,0){10}}
\put(10,10){\makebox(0,0)[cc]{\large $>$}}
\put(5,5){\makebox(0,0)[cc]{1}}
\put(15,5){\makebox(0,0)[cc]{2}}
\end{picture} \\[5pt]
\small
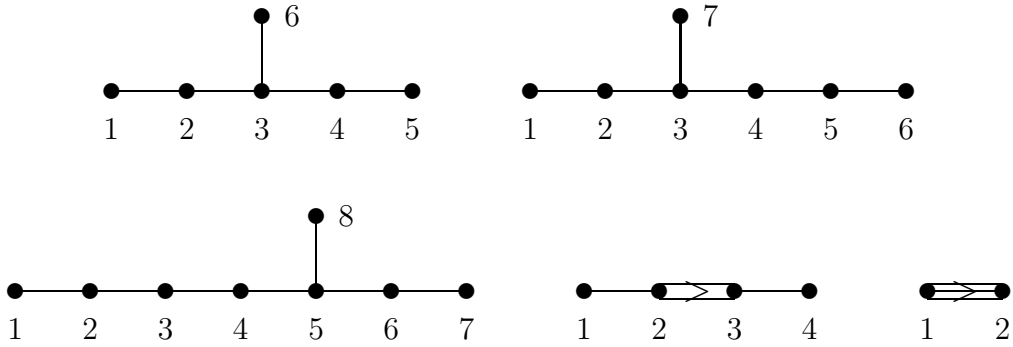
Fig.4. \it Dynkin diagrams for $E_6$, $E_7$, $E_8$, $F_4$ and $G_2$
Lie algebras respectively
\end{center}
Relation (\ref{3.4}) for these algebras takes the following form
\ber{3.18}
-\frac12\eps_s\nu_s^2(U^s,U^s)=\left\{\begin{array}{ll}
8,15,21,15,8,11 & \mbox{for }E_6, \ s=1,\dots,6; \\[5pt]
17,33,48,\ds\frac{75}2,26,\ds\frac{27}2,\ds\frac{49}2 &
\mbox{for }E_7, \ s=1,\dots,7; \\[7pt]
29,57,84,110,135,91,46,68 & \mbox{for }E_8, \ s=1,\dots,8; \\[5pt]
11,21,15,8 & \mbox{for }F_4, \ s=1,\dots,4; \\[5pt]
5,3 & \mbox{for }G_2, \ s=1,2.
\end{array}\right.
\eer

\subsection{Hyperbolic Kac-Moody algebras}

Let $\det A_{(i)}<0$. Among irreducible symmetrizable matrices satisfying
(\ref{3.6}) there exists a large subclass of Cartan matrices, corresponding
to infinite-dimensional simple hyperbolic generalized Kac-Moody (KM)
algebras of ranks $r=2,\dots,10$ \cite{Kac,FS}.

{\bf Example.} Let $A_{(i)}$ be a Cartan matrix corresponding to
$E_{10}$ hyperbolic KM algebra with the Dynkin diagram pictured
on Fig.5.
\begin{center}
\bigskip
\begin{picture}(86,21)
\put(5,10){\circle*{2}}
\put(15,10){\circle*{2}}
\put(25,10){\circle*{2}}
\put(35,10){\circle*{2}}
\put(45,10){\circle*{2}}
\put(55,10){\circle*{2}}
\put(65,10){\circle*{2}}
\put(75,10){\circle*{2}}
\put(85,10){\circle*{2}}
\put(5,10){\line(1,0){80}}
\put(65,20){\circle*{2}}
\put(65,10){\line(0,1){10}}
\put(5,5){\makebox(0,0)[cc]{1}}
\put(15,5){\makebox(0,0)[cc]{2}}
\put(25,5){\makebox(0,0)[cc]{3}}
\put(35,5){\makebox(0,0)[cc]{4}}
\put(45,5){\makebox(0,0)[cc]{5}}
\put(55,5){\makebox(0,0)[cc]{6}}
\put(65,5){\makebox(0,0)[cc]{7}}
\put(75,5){\makebox(0,0)[cc]{8}}
\put(85,5){\makebox(0,0)[cc]{9}}
\put(68,20){\makebox(0,0)[lc]{10}}
\end{picture} \\[5pt]
\small
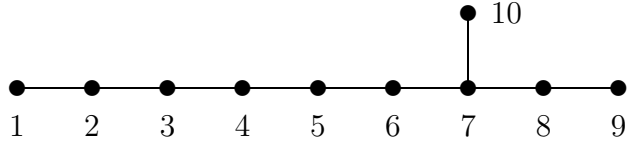
Fig.5. \it Dynkin diagram for $E_{10}$ hyperbolic KM algebra
\end{center}
Using the explicit form for $(-A_{(i)}^{-1})$ we get from (\ref{3.4})
\ber{3.21}
\frac12\eps_s(U^s,U^s)\nu_s^2=30,61,93,126,160,195,231,153,76,115
\eer
for $s=1,2,\dots,10$ respectively.

In this example for hyperbolic algebra the following relation is
satisfied:
\ber{3.22}
\eps_s(U^s,U^s) >0,
\eer
$s\in S_i$. This relation is valid in a general case, since
$(A_{(i)}^{-1})_{ss'}\le0$, $s,s'\in S$, for any hyperbolic algebra.
An example of a solution corresponding to the hyperbolic Lie algebra
${\cal F}_3$ was considered in \cite{IKM}.

\section{Solutions with intersecting $p$-branes}
\subsection{The model}

Now we consider a multidimensional gravitational model governed by
the action \cite{IMC}
\ber{4.1}
S=\int d^Dz\sqrt{|g|}\biggl\{R[g]-h_{\alpha\beta}g^{MN}\p_M\varphi^\alpha
\p_N\varphi^\beta-\sum_{a\in\tri}\frac{\theta_a}{n_a!}
\exp[2\lambda_a(\varphi)](F^a)^2\biggr\}
\eer
where $g=g_{MN}dz^M\otimes dz^N$ is the metric,
$\varphi=(\varphi^\alpha)\in\R^l$ is a vector of scalar fields,
$(h_{\alpha\beta})$ is a non-degenerate symmetric $l\times l$ matrix
$(l\in \N)$, $\theta_a=\pm1$, $F^a=dA^a$ is $n_a$-form ($n_a\ge1$),
$\lambda_a$ is a 1-form on $\R^l$: $\lambda_a(\varphi)=
\lambda_{\alpha a}\varphi^\alpha$, $a\in\tri$, $\alpha=1,\dots,l$.
Here $\tri$ is some finite set.

We consider the manifold $M=M_0\times M_1\times\dots\times M_n$, with the
metric
\ber{4.3}
g=\e{2\gamma(x)}g^0+\sum_{i=1}^n\e{2\phi^i(x)}g^i
\eer
where $g^0=g_{\mu\nu}^0(x)dx^\mu\otimes dx^\nu$ is a metric on the
manifold $M_0$, and $g^i=g_{m_in_i}^i(y_i)dy_i^{m_i}\otimes dy_i^{n_i}$
is a metric on the Ricci-flat manifold $M_i$.

Any manifold $M_\nu$ is supposed to be oriented and connected and
$d_\nu\equiv\dim M_\nu$, $\nu=0,\dots,n$. Let
\ber{4.5}
\tau_i \equiv\sqrt{|g^i(y_i)|}dy_i^1\wedge\dots\wedge dy_i^{d_i}, \quad
\eps(i)\equiv\sign(\det(g_{m_in_i}^i))=\pm1
\eer
denote the volume $d_i$-form and signature parameter respectively,
$i=1,\dots,n$. Let $\Omega=\Omega_n$ be a set of all subsets of
$\{1,\dots,n\}$, $|\Omega|=2^n$. For any $I=\{i_1,\dots,i_k\}\in\Omega$,
$i_1<\dots<i_k$, we denote
\ber{4.6}
\tau(I)\equiv\tau_{i_1}\wedge\dots\wedge\tau_{i_k},
\quad d(I)\equiv\sum_{i\in I}d_i,
\quad \eps(I) \equiv \prod_{i \in I} \eps(i).
\eer
We also put $\tau(\emptyset)= \eps(\emptyset)=
1$ and $d(\emptyset)=0$.

For fields of forms we consider the following composite electromagnetic
ansatz
\ber{4.7}
F^a=\sum_{I\in\Omega_{a,e}}{\cal F}^{(a,e,I)}+
\sum_{J\in\Omega_{a,m}}{\cal F}^{(a,m,J)}
\eer
where
\ber{4.8}
{\cal F}^{(a,e,I)}=d\Phi^{(a,e,I)}\wedge\tau(I), \mm
\label{4.9}
{\cal F}^{(a,m,J)}=\e{-2\lambda_a(\varphi)}*(d\Phi^{(a,m,J)}
\wedge\tau(J))
\eer
are elementary forms of electric and magnetic types respectively,
$a\in\tri$, $I\in\Omega_{a,e}$, $J\in\Omega_{a,m}$ and
$\Omega_{a,e}\subset\Omega$, $\Omega_{a,m}\subset\Omega$. In (\ref{4.9})
$*=*[g]$ is the Hodge operator on $(M,g)$. For scalar functions we put
$\varphi^\alpha=\varphi^\alpha(x)$, $\Phi^s=\Phi^s(x)$, $s\in S$, i.e.
they depend only on coordinates of $M_0$. Here and below
\ber{4.11}
S=S_e\sqcup S_m, \quad
S_v=\bigcup_{a\in\tri}\{a\}\times\{v\}\times\Omega_{a,v},
\eer
$v=e,m$.

Due to (\ref{4.8}) and (\ref{4.9})
\ber{4.12}
d(I)=n_a-1, \quad d(J)=D-n_a-1,
\eer
for $I\in\Omega_{a,e}$, $J\in\Omega_{a,m}$.

Let $d_0 \neq 2$ and
\ber{4.13}
\gamma=\gamma_0(\phi) \equiv
\frac1{2-d_0}\sum_{j=1}^nd_j\phi^j,
\eer
i.e. the generalized harmonic gauge is used.

Now we impose restrictions on sets $\Omega_{a,v}$. These restrictions
guarantee the block-diagonal structure of a stress-energy tensor (like
for the metric) and the existence of the $\sigma$-model representation
\cite{IMC}.

We denote $w_1\equiv\{i|i\in\{1,\dots,n\},\quad d_i=1\}$, and
$n_1=|w_1|$ (i.e. $n_1$ is the number of 1-dimensional spaces among
$M_i$, $i=1,\dots,n$).

{\bf Restriction 1.} Let 1a) $n_1\le1$ or 1b) $n_1\ge2$ and for
any $a\in\tri$, $v\in\{e,m\}$, $i,j\in w_1$, $i<j$, there are no
$I,J\in\Omega_{a,v}$ such that $i\in I$, $j\in J$ and $I\setminus\{i\}=
J\setminus\{j\}$.

{\bf Restriction 2} (only for $d_0=1,3$). Let 2a) $n_1=0$ or
2b) $n_1\ge1$ and for any $a\in\tri$, $i\in w_1$ there are no
$I\in\Omega_{a,m}$, $J\in\Omega_{a,e}$ such that $\bar I=\{i\}\sqcup J$
for $d_0 = 1$ and
$J=\{i\}\sqcup \bar I$ for $d_0 = 3$. Here and in what
follows
\ber{4.13a}
\bar I\equiv\{1,\ldots,n\}\setminus I.
\eer

It was proved in \cite{IMC} that equations of motion for the model
(\ref{4.1}) and the Bianchi identities: $d{\cal F}^s=0$, $s\in S_m$, for
fields from (\ref{4.3})--(\ref{4.13}), when Restrictions 1 and 2 are
imposed, are equivalent to equations of motion for the $\sigma$-model
(\ref{2.1}) with $(\sigma^A)=(\phi^i,\varphi^\alpha)$, the index set
$S$ from (\ref{4.11}), the target space metric $(\hat G_{AB})=\diag
(G_{ij},h_{\alpha\beta})$, with
\ber{4.15}
G_{ij}= d_i\delta_{ij}+\frac{d_i d_j}{d_0-2},
\eer
vectors
\ber{4.17}
(U_A^s)=(d_i\delta_{iI_s},-\chi_s\lambda_{\alpha a_s}),
\eer
where $s=(a_s,v_s,I_s)$, $\chi_s = +1,-1$ for $v_s = e,m$ respectively,
\ber{4.i}
\delta_{iI}= \sum_{j\in I}\delta_{ij}
\eer
is the indicator of $i$ belonging
to $I$: $\delta_{iI}=1$ for $i\in I$ and $\delta_{iI}=0$ otherwise; and
\ber{4.18}
\eps_s=(-\eps[g])^{(1-\chi_s)/2}\eps(I_s) \theta_{a_s},
\eer
$s\in S$, $\eps[g]\equiv\sign\det(g_{MN})$.

{\bf General intersection rules.} Scalar products (\ref{2.2}) for
vectors $U^s$  were calculated in \cite{IMC}
\ber{4.19}
(U^s,U^{s'})=d(I_s\cap I_{s'})+\frac{d(I_s)d(I_{s'})}{2-D}+
\chi_s\chi_{s'}\lambda_{\alpha a_s}\lambda_{\beta a_{s'}}h^{\alpha\beta}
\equiv B^{ss'},
\eer
where $(h^{\alpha\beta})=(h_{\alpha\beta})^{-1}$; $s=(a_s,v_s,I_s)$ and
$s'=(a_{s'},v_{s'},I_{s'})$ belong to $S$. Let us put $(U^s,U^s)\ne0$,
$s \in S$. Then, we obtain the general intersection rule formulas
\ber{5.8}
d(I_s\cap I_{s'})=\frac{d(I_s)d(I_{s'})}{D-2}-
\chi_s\chi_{s'}\lambda_{\alpha a_s}\lambda_{\beta a_{s'}}h^{\alpha\beta}
+\frac12(U^{s'},U^{s'})A^{ss'},
\eer
$s\ne s'$, where $(A^{ss'})$ is the quasi-Cartan matrix (\ref{3.2})
(see also (6.32) from \cite{IMJ}).

\subsection{Exact solutions}

Applying the results from Sect. 2 to our multidimensional model 
(\ref{4.1}), when vectors $(U^s,s\in S)$ obey the block-orthogonal 
decomposition (\ref{2.3}), (\ref{2.4}) with scalar products defined in 
(\ref{4.19}) we get the exact solution:
\ber{4.20}
g=U\left\{g^0+\sum_{i=1}^n U_ig^i\right\}, \mm
\label{4.21}
U=\left(\prod_{s\in S}H_s^{2d(I_s)\eps_s\nu_s^2}\right)^{1/(2-D)}, \quad
U_i=\prod_{s\in S}H_s^{2\eps_s\nu_s^2\delta_{iI_s}}, \mm
\label{4.23}
\varphi^\alpha=-\sum_{s\in S}\lambda_{a_s}^\alpha\chi_s
\eps_s\nu_s^2\ln H_s, \quad
F^a=\sum_{s\in S}{\cal F}^s\delta_{a_s}^a,
\eer
where
\ber{4.25}
{\cal F}^s=\nu_sdH_s^{-1}\wedge\tau(I_s), \mbox{ for } v_s=e, \mm
\label{4.26}
{\cal F}^s=\nu_s(*_0dH_s)\wedge\tau(\bar I_s), \mbox{ for } v_s=m,
\eer
$H_s$ are harmonic functions on $(M_0,g^0)$ coinciding inside blocks
of matrix $(B^{ss'})$ from (\ref{4.19}) ($H_s=H_{s'}$, $s,s'\in S_j$,
$j=1,\dots,k$, see (\ref{2.3}), (\ref{2.4})) and relations
\ber{4.27}
\sum_{s'\in S} B^{ss'} \eps_{s'}\nu_{s'}^2=-1
\eer
on the matrix $(B^{ss'})$,
parameters $\eps_s$ (\ref{4.18}) and  $\nu_s$ are imposed, $s\in S$,
$i=1,\dots,n$; $\alpha=1,\dots,l$. Here $\lambda_a^\alpha=
h^{\alpha\beta}\lambda_{\beta a}$, $*_0=*[g^0]$ is the Hodge operator
on $(M_0,g^0)$ and $\bar I$  is defined in (\ref{4.13a}).

In deriving the solution we used as in \cite{IMC} the relations for
contravariant components of $U^s$-vectors:
\ber{4.29a}
U^{si}=\delta_{iI_s}-\frac{d(I_s)}{D-2}, \quad
U^{s\alpha}=-\chi_s\lambda_{a_s}^\alpha,
\eer
$s=(a_s,v_s,I_s)$.

{\bf Remark 1.} The solution is also valid for $d_0=2$, if Restriction 2
is replaced by Restriction $2^{*}$ below. It may be proved using a more
general form of the sigma-model representation (see Remark 2 in \cite{IMC}).

{\bf Restriction 2${}^*$} (for $d_0=2$). For any  $a \in \tri$  there are
no $I\in\Omega_{a,m}$, $J\in\Omega_{a,e}$ such that $\bar I =  J$ and for
$n_1 \geq 2$, $i,j\in w_1$, $i \neq j$, there are no $I\in\Omega_{a,m}$,
$J\in\Omega_{a,e}$ such that $i\in I$, $j\in \bar J$, $I\setminus\{i\}=
\bar J \setminus \{j\}$.

{\bf Remark 2.} Our terminology: "orthogonal", "block-orthogonal"
(intersections), refers to the scalar products of $U$-vectors (\ref{2.2})
but not to the orientation of $p$-branes in the multidimensional space-time
with the metric (\ref{4.3}). The metric (\ref{4.3}) is block-diagonal, hence
the $p$-branes are orthogonal or parallel with respect to the metric
(\ref{4.3}) even if their intersections rules are "non-orthogonal" ones.
(Here we do not consider "branes at angles").

\section{Some examples}

Here we consider several examples of the solutions corresponding to the Lie
algebra $A_2$.

\subsection{IIA supergravity}

Let us consider the $D=10$ IIA supergravity with the bosonic part of the
action
\ber{6.1}
S=\int d^{10}z\sqrt{|g|}\biggl\{R[g]-(\p\varphi)^2-\sum_{a=2}^4
\e{2\lambda_a\varphi}F_a^2\biggr\}-\frac12\int F_4\wedge F_4\wedge A_2,
\eer
where $F_a=dA_{a-1}+\delta_{a4}A_1\wedge F_3$ is an $a$-form and
\ber{6.2}
\lambda_3=-2\lambda_4, \quad \lambda_2=3\lambda_4, \quad
\lambda_4^2=\frac18.
\eer
The dimensions of $p$-brane worldsheets are
\ber{6.3}
d(I)=\left\{\begin{array}{ll}
1,2,3&\mbox{electric case} \\
7,6,5&\mbox{magnetic case}
\end{array}\right.
\eer
for $a=2,3,4$ respectively.

We consider here the sector corresponding to $a=3,4$ describing electric
$p$-branes: fundamental string (FS), D2-brane and magnetic $p$-branes:
NS5- and D4-branes.

We get $(U^s,U^s)=2$ for all $s$. The solutions with $A_2$ intersection
rules corresponding to relations
\ber{6.4}
1 \cap 5=2 \cap 5=2 \cap 4=1 \mm
5\cap4=5\cap5=3, \quad 4\cap 4= 2
\eer
are valid in the "truncated case" (without Chern-Simons terms) and in
a general case (\ref{6.1}) as well. Here $(p_1 \cap p_2= d)\Leftrightarrow
(d(I)=p_1 + 1, d(J)= p_2 + 1, d(I\cap J) = d)$.

Let us consider the solution describing the electromagnetic dyon with
intersections
\ber{6.5}
1\cap5=2\cap4=1.
\eer
Here the solution reads
\ber{6.6}
g=H^2g^0-H^{-2}dt\otimes dt+g^1+g^2, \mm
\label{6.7}
F_a=\nu_1dH^{-1}\wedge dt\wedge\tau_1+\nu_2(*_0dH)\wedge\tau_1,
\eer
where $H$ is harmonic function on $(M_0,g^0)$, $d_0=3$, $\nu_1^2=\nu_2^2=1$,
$d_1=a-2$, $d_2=8-a$, $a=3,4$. The signature restrictions are the following:
$\eps_1=+1$, $\eps_2=-\eps[g]=+1$.

For $g^0 = \sum_{\mu =1}^{3} dx^{\mu} \otimes dx^{\mu}$ and
\ber{6.7a}
H=1+\sum_{i=1}^{N} \frac{q_i}{|x- x_i|},
\eer
the  4-dimensional section of the metric (\ref{6.6}) coincides with the
standard Majumdar-Papapetrou solution \cite{MP} describing $N$ extremal
charged black holes with horizons at points $x_i$, and charges $q_i>0$,
$i=1,\ldots,N$. Our solution (\ref{6.6})--(\ref{6.7}) with $H$ from
(\ref{6.7a}) describes $N$ extremal $p$-brane black holes formed by dyons.
Any dyon contains one electric "brane" and one magnetic "brane" with equal
charge densities.

\subsection{$D=11$ supergravity}

For the $D=11$ supergravity with the bosonic part of the action
\ber{6.8}
S=\int d^{11}z\sqrt{|g|}\biggl\{R[g]-\frac{F_4^2}{4!}\biggr\}+
c\int F_4\wedge F_4\wedge A_3
\eer
we get the $A_2$-solutions with the intersections
\ber{6.9}
2\cap5=1, \quad 5\cap5=3.
\eer

The electromagnetic dyon with the intersection $2\cap5=1$ is given by
relations (\ref{6.6}) and (\ref{6.7}) with $a=4$, $d_1=3$, $d_2=5$. Other
relations are unchanged.

\section{Discussions}

Here we obtained explicit formulas for the solution from \cite{IM5}
corresponding to direct sums of finite-dimensional simple Lie algebras. These
solutions for $p$-brane case have nonstandard intersection rules, defined
by Cartan matrices of the corresponding Lie algebras and describe bound
state configurations of $p$-branes.

Here we presented several examples of such dyons corresponding to the
$A_2$ Lie algebra in $D=10$ IIA and $D=11$ supergravities. We note that
the usual solutions in IIA supergravity corresponding to the Lie
algebras $A_1\oplus A_1$ (D4- and NS5-branes) play a rather important
role in MQCD \cite{EGK}--\cite{BIKSY}. It is also interesting to consider
the possible applications of the obtained solutions in MQCD.

\begin{center}
{\bf Acknowledgments}
\end{center}

This work was supported in part by the DFG grant 436 RUS 113/236/O(R),
by the Russian Ministry of Science and Technology and by RFBR grant
98-02-16414. One of us (V.D.I.) thanks H. Nicolai for the hospitality
during his visit to Albert-Einstein-Institut (Max-Planck-Institut of
Gravitational Physik, Potsdam).


\small
\begin{thebibliography}{99}

\bibitem{MP}
S.D. Majumdar, {\it Phys. Rev. } {\bf 72}, 930 (1947);  \\
A. Papapetrou,  {\it Proc. R. Irish Acad. } {\bf A51}, 191 (1947).

\bibitem{IM5}
V.D. Ivashchuk and V.N. Melnikov, Majumdar-Papapetrou Type Solutions
in Sigma-model and Intersecting p-branes, hep-th/9802121, {\it submitted
to Class. and Quant. Grav.}

\bibitem{IM4}
V.D. Ivashchuk and V.N. Melnikov,
Intersecting p-brane Solutions in Multidimensional Gravity and M-theory,
{\it Gravitation and Cosmology } {\bf 2}, No 4, 204 (1996); hep-th/9612089.

\bibitem{IM}
V.D. Ivashchuk and V.N. Melnikov, {\it Phys. Lett.} {\bf B 403 }
23 (1997).

\bibitem{IMC}
V.D. Ivashchuk and V.N. Melnikov, Sigma-model for the Generalized
Composite p-branes, {\it Class. Quant. Grav.} 14 (11), 3001 (1997);
hep-th/9705036.

\bibitem{IMR}
V.D. Ivashchuk, V.N. Melnikov and M. Rainer, Multidimensional
$\sigma$-models with Composite Electric p-branes, gr-qc/9705005;
{\it Gravitation and Cosmology} {\bf 4}, No. 1 (13), 73 (1998).

\bibitem{S}
K.S. Stelle, Lectures on Supergravity p-branes, hep-th/9701088.

\bibitem{PT}
G. Papadopoulos and P.K. Townsend, {\it  Phys. Lett.} {\bf B 380 }
273 (1996).

\bibitem{Ts}
A.A. Tseytlin, {\it Nucl. Phys.} {\bf B 475 } 149 (1996); hep-th/9604035.

\bibitem{GKT}
J.P. Gauntlett, D.A. Kastor, and J. Traschen, {\it Nucl. Phys.}
{\bf B 478 } 544 (1996); hep-th/9604179.

\bibitem{AV}
I.Ya. Aref'eva and A.I. Volovich, {\it Class. Quant. Grav.} 14 (11),
2990 (1997); hep-th/9611026.

\bibitem{AR}
I.Ya. Aref'eva and O.A. Rytchkov, Incidence Matrix Description of
Intersecting p-brane Solutions, {\it Preprint } SMI-25-96; hep-th/9612236.

\bibitem{AEH}
R. Argurio, F. Englert and L. Hourant, {\it Phys. Lett.} {B 398 }
2991 (1997); hep-th/9701042.

\bibitem{EGK}
S. Elitzur, A. Giveon and D. Kutasov, Branes and $N=1$ Duality in String
Theory, {\it Phys. Lett.} {\bf B400 } (1997) 269.

\bibitem{BH}
J. Brodie and A. Hanany, Type IIA Superstrings, Chiral Symmetry, and
$N=1$ 4D Gauge Theory Dualities, {\it Nucl. Phys.} {\bf B506 } (1997) 157;
hep-th/9704043.

\bibitem{BSTY}
A. Brandhuber, J. Sonnenschein, S. Theisen and S. Yankielowicz,
Brane Configurations and 4D Field Theory Dualities, {\it Nucl. Phys.}
{\bf B502 } (1997) 125; hep-th/9704044.

\bibitem{Witten}
E. Witten, Branes and the Dynamics of QCD, {\it Nucl. Phys.} {\bf B507 }
(1997) 658; hep-th/9706109.

\bibitem{HOO}
K. Hori, H. Ooguri and Y. Oz, Strong Coupling Dynamics of
Four-Dimensional $N=1$ Gauge Theory from M-Theory Five-brane,
hep-th/9706082.

\bibitem{HSZ}
A. Hanany, M. Strassler and A. Zaffaroni, Confinement and Strings in MQCD,
hep-th/9707244.

\bibitem{HOO2}
Jan de Boer, Kentaro Hori, Hirosi Ooguri and Yaron Oz, Kahler Potential
and Higher Derivative Terms from M-Theory Five-brane, hep-th/9711143.

\bibitem{BIKSY}
A. Brandhuber, N. Itzaki. V. Kaplunovsky, J. Sonnenschein
and S. Yankielowicz, Comments on the M-Theory Approach to $N=1$ SQCD and
Brane Dynamics, {\it Phys. Lett.} {\bf 410 } (1997) 27; hep-th/9706127.

\bibitem{IMJ}
V.D. Ivashchuk and V.N. Melnikov, Multidimensional Classical and
Quantum Cosmology with Intersecting $p$-branes, hep-th/9708157;
{\it J. Math. Phys.}, {\bf 39}, 2866 (1998).

\bibitem{Br}
K.A. Bronnikov, Block-orthogonal Brane systems, Black Holes and
Wormholes, hep-th/9710207;
{\it Gravitation and Cosmology} {\bf 4}, No 1(13), 49 (1998).

\bibitem{Kac}
V.G. Kac, Infinite-dimensional Lie Algebras (Cambridge University
Press, Cambridge, 1990).

\bibitem{FS}
J. Fuchs and C. Schweigert, Symmetries, Lie algebras and Representations.
(Cambridge University Press, Cambridge, 1997).

\bibitem{IKM}
V.D. Ivashchuk, S.-W. Kim and V.N. Melnikov, Hyperbolic Kac-Moody Algebra
from Intersecting p-Branes, hep-th/9803006.

\end{thebibliography}
\end{document}